\documentstyle[12pt,axodraw]{article}
\newcommand{\nn}{\nonumber}
\newcommand{\bd}{\begin{document}}
\newcommand{\ed}{\end{document}}
\newcommand{\bc}{\begin{center}}
\newcommand{\ec}{\end{center}}
\newcommand{\be}{\begin{eqnarray}}
\newcommand{\ee}{\end{eqnarray}}
\newcommand{\ba}{\begin{array}}
\newcommand{\ea}{\end{array}}
\newcommand{\strich}[1]{#1  \! \! \slash}
\newcommand{\eqn}{\global\def\theequation}
\newcommand{\sw}{sin^2 \theta_W}
\newcommand{\fbd}{f_B}
\renewcommand{\thefootnote}{\alph{footnote}}
\newcommand{\se}{\section}
\newcommand{\sse}{\subsection}
\newcommand{\bi}{\bibitem}
\def\figcap{\section*{Figure Captions\markboth
     {FIGURECAPTIONS}{FIGURECAPTIONS}}\list

     {Figure \arabic{enumi}:\hfill}{\settowidth\labelwidth{Figure 999:}
     \leftmargin\labelwidth
     \advance\leftmargin\labelsep\usecounter{enumi}}}
\let\endfigcap\endlist \relax
\def\reflist{\section*{References\markboth
     {REFLIST}{REFLIST}}\list
     {[\arabic{enumi}]\hfill}{\settowidth\labelwidth{[999]}

     \leftmargin\labelwidth
     \advance\leftmargin\labelsep\usecounter{enumi}}}
\let\endreflist\endlist \relax

\begin{document}
\tolerance=10000
\baselineskip=7mm
\begin{titlepage}

 \vskip 0.5in
 \null
\begin{center}
 \vspace{.15in}
{\LARGE {\bf Tau Radiative Decays in the Light Front Quark Model} }\\
\vspace{1.0cm}
%\begin{center}
  \par
 \vskip 2.1em
 {\large
  \begin{tabular}[t]{c}
{\bf C.~Q.~Geng and C.~C.~Lih}
\\
\\
{\sl ${}$Department of Physics, National Tsing Hua University}
\\  {\sl  $\ $ Hsinchu, Taiwan, Republic of China }
\\
\\
   \end{tabular}}
 \par \vskip 5.3em
 {\Large\bf Abstract}
\end{center}

We  study the decays of $\tau^- \to \nu_{\tau} P^-\gamma\
(P=\pi^-\,,K^-)$ in the light front quark model. We calculate the
form factors and use them to evaluate the decay widths.
 We find that, in the standard
model, the decay widths are $1.62\times 10^{-2}\,\, (3.86\times
10^{-3})\, \Gamma_{\tau^- \to \nu_{\tau} \pi^- }$ and $1.91\times
10^{-3}\,\,(5.38\times 10^{-4})\, \Gamma_{\tau^- \to \nu_{\tau}
K^- }$ with the cuts of $E_{\gamma}=50\,\,(400)$ MeV and $t_{0}=
800\,\,(1200)$ MeV for $\tau^- \to \nu_{\tau} \pi^- \gamma$ and
$\tau^- \to \nu_{\tau} K^- \gamma$, respectively. We also show
that with including the radiative decay widths, the experimental
rate for $\tau^-\to \nu_\tau P^-$ can be explained.

\end{titlepage}

\se{Introduction}

\ \ \ Tau is the only charged lepton which decays into hadrons.
Theoretically, the hadronic $\tau$ decays can provide us with
valuable information on strong interaction. The simplest rare
radiative $\tau$ decays are $\tau^- \to \nu_{\tau} P^{-} \gamma$ with
$P$ being the pseudoscalar mesons of $\pi$ and $K$.
% which are analogous to the radiative $P$ decays. However, unlike
%the radiative meson decays, the monemtum transfer squared $p^2$ in
%the $\tau$ decays can take any values up to $m^{2}_{\tau}$ .
The contributions to the decays can be divided into
``internal-bremsstrahlung'' (IB) and  ``structure-dependent'' (SD)
parts in terms of the photon emission. For the IB contribution the
photon  emits from the $\tau$ lepton as well as external hadron,
while the SD one from intermediate states
%governed by strong interaction and
described by the vector and axial-vector form factors, $F_{V,A}$. Our
main task is to calculate these form factors which are functions
of $t^2=(p+q)^2$ where $q\; (p)$ is the four momentum of $\gamma\;
(P)$ and $M^{2}_{P} \leq t^2 \leq m_{\tau}^{2}$. In general, the
momentum  dependences of $F_{V,A}$ could also help us to determine
the bound state wave functions of mesons.

In this paper, we will use the light front quark model (LFQM) to
evaluate the matrix elements in $\tau\to\nu_{\tau}P \gamma$
decays. The LFQM has been widely applied to study the form factors
of weak decays \cite{jau,Don94,Don95,cheng1,cheng2,geng1,geng2}.
It is the relativistic quark model \cite{wmz} in which a
consistent and relativistic treatment of quark spins and the
center-of-mass motion can be carried out. Moreover, the meson
state of the definite spins could be relativistically constructed
by the Melosh transformation \cite{melosh}. They are many
advantages in the LFQM. For example, the light front wave function
is manifestly Lorentz invariant as it is expressed in terms of the
momentum fraction variable (in ``+'' component) in analogy with
the parton distributions in the infinite momentum frame. The
kinematic subgroup of the light front formalism has the maximum
number of interaction-free generators including the boost operator
which describes the center-of-mass motion of the bound state
\cite{wmz}.

The paper is organized as follows. In Sec.~2, we present the
matrix elements and study the form factors in the $P \to \gamma$
transitions within the framework of the LFQM. We calculate the
decay widths of $\tau \to \nu_{\tau} P \gamma$ in Sec.~3. We also
compare our results with those in literature \cite{tau2,tau3}. We
give our conclusions in Sec.~4.

\se{Form Factors in the Light Front Quark Model}

\ \ \ Similar to the radiative meson decay, the decay amplitude
for \be
\tau^{-}(l) &\to &\nu_{\tau}(k) P^{-}(p) \gamma(q) %
\ee with $P=\pi$ or $K$  can be written as  \cite{tau2,tau1,pi}
\be {\cal M} &=&{\cal M}_{IB}+{\cal M}_{SD}\,,
 \label{amp1}
 \\
{\cal M}_{IB} &=& G_{F} \cos \theta_{c} e f_{P} m_{\tau}
\bar{u}(k) (1+\gamma_{5}) \left[\frac{p \cdot \epsilon}{p \cdot
q}+\frac{\strich{q} \strich{\epsilon}}{2 l \cdot q}-\frac{l \cdot
\epsilon}{l \cdot q} \right] v(l)\,,%
 \nonumber \\ %
 {\cal M}_{SD} &=&
\frac{G_{F} \cos \theta_{c} e}{\sqrt{2}} \Bigg\{ i \epsilon_{\mu
\nu \rho \sigma} L^{\mu} \epsilon^{\nu} q^{\rho} p^{\sigma}
\frac{F_{V}}{M_{P}} \nonumber \\ &&+\bar{u}(k) (1+\gamma_{5})
\left[(p\cdot q) \strich{\epsilon}- (\epsilon \cdot p) \strich{q}
\right] v(l) \frac{F_{A}}{M_{P}} \Bigg\} \,,%
\label{amp2} %
\ee %
where $L^{\mu}=\bar{u}(k)\gamma^{\mu}(1-\gamma_{5})u(l)$ and
$F_{A,V}$ are the form-factors corresponding to the vector and
axial-vector currents, defined by
\be
\langle\gamma (q),P(p)|\,\bar{q}\gamma_{\mu }\gamma _{5}b|\,0\,
\rangle &=& e{\frac{F_{A}}{M_{P}}}\left[ (p\cdot q) \epsilon^*_\mu
- (\epsilon ^{*}\cdot p)q_{\mu }  \right] , \nonumber\\
\langle\gamma(q),P(p)|\,\bar{q}\gamma_{\mu}b|\,0\, \rangle &=&
ie{\frac{F_{V}}{M_{P}}}\varepsilon_{
\mu\nu\alpha\beta}\epsilon^{*\nu} q^\alpha p^\beta , \label{ff}
\ee %
with  $\epsilon _{\mu }$ being the photon polarization vector and
$q$ ($p$) the four momentum of $\gamma\,(P)$, respectively. The
form factors $F_{A,V}$ in Eq. (\ref{ff}) depend on
$t^{2}=(p+q)^{2}$ for which the allowed range is $M_{P}^{2} \leq
t^{2} \leq m_{\tau}^{2}$, in contrast to $0 \leq t^{2} \leq
M_{P}^{2}$ in $P^+ \to l^+ \nu_{l}\gamma$.

\sse{Light front formalism}

To calculate of the hadronic matrix elements in Eq.~(\ref{ff}),
one usually lets $t^+=0$ to have a spacelike momentum transfer.
However, since the momentum transfer should be always timelike in
a real decay process, in this work we use the frame of $t_\perp=0$
with the physically accessible kinematic region of $0\leq
t^{2}\leq t^2_{\max }$. Within the light front formalism, the
meson bound state, which consists of a quark $q_1$ and an
anti-quark $\bar{q}_2$ with the total momentum $p$ and spin $S$,
can be written as
\begin{eqnarray}
|M(p,S,S_z)\rangle &=& \int [dk_{1}][dk_{2}] 2(2\pi)^{3}\delta
^{3}(p-k_{1}-k_{2}) \nonumber \\ &\times&\sum_{\lambda
_{1}\lambda_{2}} \Phi^{S S_{z}}(k_1,k_2,\lambda_1,\lambda_2)
b_{q_{1}}^{+}(k_{1},\lambda _{1}) d_{\bar{q_{2}}}^{+}(
k_{2},\lambda _{2}) |\,0\,\rangle\,, \label{mwf}
\end{eqnarray}
where
\be
&&[dk]={\frac{dk^+dk_{\bot}}{2(2\pi)^3 k^{+}}}\,,  \nonumber \\
&&\{b_{\lambda'}(k'),b^\dagger_{\lambda}(k)\}
=\{d_{\lambda'}(k'),d^\dagger_{\lambda}(k)\}
=2(2\pi)^3~\delta^3(k'-k)~\delta_{\lambda'\lambda}\,,
 \ee
  and
$k_{1(2)}$ is the on-mass shell light front momentum of the
internal quark $q_{1}(\bar{q}_2)$. The light front relative
momentum variables $(x,k_{\bot})$ are defined by
\begin{eqnarray}
        && k^+_1=x_1 p^+, \quad k^+_2=x_2 p^+, \quad x_1+x_2=1, \nonumber \\
        && k_{1\bot}=x_1 p_\bot+k_\bot, \quad k_{2\bot}=x_2
                p_\bot-k_\bot\,.
\end{eqnarray}
In the momentum space, the wave function $\Phi^{SS_z}$ is given by
\begin{equation}
        \Phi^{SS_z}(k_1,k_2,\lambda_1,\lambda_2)
                = R^{SS_z}_{\lambda_1\lambda_2}(x,k_\bot)~ \phi(x, k_\bot),
                \label{phi1}
\end{equation}
where $\phi(x,k_{\bot})$ describes the momentum distribution
amplitude of the constituents in the bound state and
$R^{SS_z}_{\lambda_1\lambda_2}$ constructs a spin state $(S,S_z)$
out of light front helicity eigenstates $(\lambda_1\lambda_2)$,
expressed by
\be
R^{SS_z}_{\lambda_1 \lambda_2}(x,k_\bot)
                =\sum_{s_1,s_2} \langle \lambda_1|
                {\cal R}_M^\dagger(1-x,k_\bot, m_1)|s_1\rangle
                \langle \lambda_2|{\cal R}_M^\dagger(x,-k_\bot, m_2)
                |s_2\rangle
                \langle {1\over2}s_1
                {1\over2}s_2|SS_z\rangle\,.
\label{n6} \ee %
In Eq. (\ref{n6}), $|s_i\rangle$ are the Pauli spinors and ${\cal
R}_M$ is the Melosh transformation operator, given by
\be
        {\cal R}_M (x,k_\bot,m_i) =
                {m_i+x_iM_0+i\vec \sigma\cdot\vec k_\bot \times \vec n
                \over \sqrt{(m_i+x_i M_0)^2 + k_\bot^2}}
\ee %
with
\be
M_0^2&=&{ m_1^2+k_\bot^2\over x_1}+{ m_2^2+k_\bot^2\over x_2}\,,
        \nn\\
n&=&(0,~0,~1) \,.%
\label{M0} %
\ee%
We note that  Eq.~(\ref{n6}) can, in fact, be expressed as a
covariant form \cite{jau}
 \be
 R^{SS_z}_{\lambda_1\lambda_2}(x,k_\bot)
                ={\sqrt{k_1^+k_2^+}\over \sqrt{2} ~{\widetilde M_0}}
        ~\bar u(k_1,\lambda_1)\Gamma v(k_2,\lambda_2),,
\label{rs}
\ee
where
\be
&&{\widetilde M_0} \equiv \sqrt{M_0^2-(m_1-m_2)^2}, \nonumber\\
&& \sum_\lambda u(k,\lambda) \overline{u}(k,\lambda) = {\frac{m + \not{\!
}k }{k^+}} \, , \ \ \sum_\lambda v(k,\lambda) \overline{v}(k,\lambda) = - {%
\frac{m - \not{\! }k }{k^+}} \, , \ee
and
\be
&&\Gamma=\gamma_5
\qquad ({\rm pseudoscalar}, S=0),
\nonumber\\
&&\Gamma=-\not{\! \hat{\varepsilon}}(S_z)+
          {\hat{\varepsilon}\cdot(k_1-k_2)
                \over M_0+m_1+m_2} \qquad ({\rm vector}, S=1),
\ee
with
\begin{eqnarray}
        &&\hat{\varepsilon}^\mu(\pm 1) =
                \left[{2\over p^+} \vec \varepsilon_\bot (\pm 1) \cdot
                \vec p_\bot,\,0,\,\vec \varepsilon_\bot (\pm 1)\right],
                \quad \vec \varepsilon_\bot
                (\pm 1)=\mp(1,\pm i)/\sqrt{2}, \nonumber\\
        &&\hat{\varepsilon}^\mu(0)={1\over M_0}\left({-M_0^2+p_\bot^2\over
                p^+},p^+,p_\bot\right).   \label{polcom}
\end{eqnarray}
The normalization condition of the meson state is given by
\be
&&\langle M(p',S',S'_z)|M(p,S,S_z)\rangle = 2(2\pi)^{3} p^{+}
\delta^{3}(\tilde p'- \tilde p)\delta_{S'S}\delta_{S'_zS_z} \, ,
\ee which leads to
\be
\int {dx\, d^2k_\bot\over 2(2\pi)^3} |\phi(x,k_\bot)|^2 = 1\, .
\ee

In principle, the momentum distribution amplitude $\phi(x,k_\bot)$
can be obtained by solving the light front QCD bound state
equation \cite{wmz2,wmz3}. However, before such first-principle
solutions are available, we would have to be contented with
phenomenological amplitudes. One example that has been often used
in the literature is the Gaussian type wave function:
\be
\phi(x,k_{\bot})=N\sqrt{\frac{dk_{z}}{dx}} \exp \left(
-\frac{\vec{k}^{2}} {2\omega_{M}^{2}}\right) \,, \label{7} \ee
where $N = 4 ( \pi/\omega_{M}^{2})^\frac{3}{4}$, $\vec k =
(k_{\bot}, k_z)$, and $dk_z/ dx = e_1 e_2/ x(1-x)M_0$ with $k_z$
being defined through
\be
x = {e_1-k_z\over e_1 + e_2} \, \ \
1-x = {e_2+k_z \over e_1 + e_2}\, , \ \  e_i = \sqrt{m_i^2 + \vec k^2} \,
\ee
by
\be
& &
\ \ k_{z} =\left( x-\frac{1}{2}\right) M_{0}+\frac{m_{1}^{2}-m_{2}^{2}}{%
2M_{0}} \,.
 \ee
 % and the Jacobian of the transformation from $(x,k_{\bot})$
%to $\vec k$.
In particular, with appropriate parameters, the wave
function in Eq. (\ref{7}) describes satisfactorily the pion
elastic form factor up to $t^2\sim 10~{\rm GeV}^2$ \cite{Chung}.

\sse{The Form Factors of $F_A$ and $F_V$}

The one-loop diagrams that contribute to $F_{V,A}$ are illustrated
in Figure 1. From the figures, the hadronic matrix elements in Eq.
(\ref{ff}) are found to be
\be
&&\langle\gamma (q),P(p )|\bar{q}\gamma_{\mu }\,(1-\gamma
_{5})\,Q|\,0\, \rangle=\int \frac{d^{4}k_1'}{(2 \pi)^{4}}
\Lambda_{P} \nonumber \\ &&\times\bigg\{\gamma_{\mu }(1-\gamma
_{5})\frac{i(\strich{k}'_{1}+m_{Q})}
{k_{1}^{'2}-m_{Q}^{2}+i\epsilon}
\gamma_{5}\frac{i(-\strich{k}^{'}_{2}+m_{q})}
{k_{2}^{'2}-m_{q}^{2}+i\epsilon} ie_{q}\strich{\epsilon}
\frac{i(\strich{k}_{1}+m_{q})} {k_{1}^{2}-m_{q}^{2}+i\epsilon}
\nonumber \\ &&-(q\leftrightarrow Q \, , k'_{1} \to k'_{2}
)\bigg\} \label{int} \ee where $\Lambda_{P}$ is a vertex function
related to $Q\bar{q}$ bound state of the meson $P$, $k_2=p-k'_1$
and $k_1=t-k'_1=k_2+q$. After integrating over the LF momentum
$k_{1}^{-}$ in Eq. (\ref{int}), we get
\be
&&\langle\gamma (q_{\gamma}),P(p
)|\bar{q}\gamma_{\mu }\,(1-\gamma _{5})\,Q|\,0\,
\rangle\nonumber \\
&&
=\int_{0}^{p} [d^{3}k'_{1}]\bigg\{\frac{1}{k_{1}^{-}-k_{1on}^{-}}
(I^{\mu\nu}|_{k_{1on}^{'-}}) \frac{\Lambda_{P}}{k_{2}^{'-}-k_{2on}^{'-}}
-(q \leftrightarrow Q\, , k'_{1} \to k'_{2} )\bigg\}
\,\,,
\label{22}
\ee
where
\be
&&[d^{3}k'_{1}]=\frac{dk_{1}^{+}dk_{1\bot}}{2(2\pi)^{3} k_1^{'+}
k_2^{'+} k_1^{+}} ~, \nonumber \\
&&I^{\mu\nu}|_{k_{1on}^{-}}=Tr\bigg\{{\gamma_{\mu } (1-\gamma
_{5})(\strich{k}^{'}_{1}+m_{Q})}
\gamma_{5}(-\strich{k}^{'}_{2}+m_{q})ie_{q}\strich{\epsilon}
(\strich{k}_{1}+m_{q})\bigg\}~, \nonumber \\
&&k_{ion}^{-}=\frac{m_{i}^{2}+k_{i\bot}^{2}}{k_{i}^{+}}~,~
k_{1(2)}^{'-}=p_{on}^{-}-k_{2(1)on}^{'-} ~,~
k_{1}^{-}=k_{2on}^{-}+q_{\gamma on}^{-}
 \,, \label{trace}%
 \ee
with $\{on\}$ representing the on-shell particles.
 By considering the
``good'' component $\mu=+$, the meson wave function in Eq.
(\ref{phi1}) and Melosh transformation in Eq. (\ref{rs}) are
related to the bound state vertex function $\Lambda_{P}$, given by
\cite{jau,dem}:
\be
\frac{\Lambda_{P}}{k_{2}^{'-}-k_{2on}^{'-}}\to
{\sqrt{k_1^{'+} k_2^{'+}}\over \sqrt{2} ~{\widetilde M_0}}~ \phi(x,
k_\bot)\,.
\ee %
Moreover, the matrix elements in Eq. (\ref{ff}) become
\be
\langle\gamma (q),P(p
)|\bar{q}\,\gamma^{+} \gamma _{5}\,Q|\,0\,
\rangle&=&e\frac{F_{A}}{2M_{P}}%
( \epsilon _{\bot }^{*}\cdot {q}_{\bot }) \,, \nonumber \\
\langle\gamma (q),P(p )|\bar{q}\,\gamma^{+}\,Q|\,0\,
\rangle&=&-ie\frac{F_{V}}{2M_{P}} \epsilon ^{ij}\epsilon _{i}^{*}{q}_{j}\,\,.
\label{ff2}%
 \ee
  Here we have used the LF monentum variables
$(x,k_{\bot})$ and worked in the frame that the transverse
momentum is purely longitudinal, $i.e.$, $t_{\bot}$ = $0$. We note
that $t^{2}=t^{+}t^{-} \geq 0$ covers the entire range of momentum
transfers. Thus, the relevant quark variables for Figure.~1 are
\be
&&k_{1}^{+}=(1+x)q^+,~~k_{2}^{+}=xq^+,
~~k_{1\perp}=(1+x)q_\perp-k_\perp,~~k_{2\perp}=xq_\perp-k_\perp\,.
\nonumber \\
&&~k_{1}^{'+}=(1-x')p^+,
~k_{2}^{'+}=x'p^+,
~k_{1\perp}^{'}=-(1-x')q_\perp+k_\perp^{'},~k_{2\perp}^{'}=-x'q_\perp-k^{'}_\perp\,,
\label{transmom}
\ee
where $x'~(x)$ is the momentum fraction of the anti-quark
in the meson (photon) state. At the quark loop, it requires that
\be
k_{2(1)}^{'+}=k_{2}^{+},~~~~~k^{'}_{2(1)\perp}=k_{2\perp}\,,
\label{momeq}%
 \ee
 for Figure~1a(b). Therefore, the trace
$I^{\mu\nu}$ in Eq.~(\ref{trace}) can be easily carried out. The
form factors $F_{A}$ and $F_{V}$ in Eq.~(\ref{ff2}) are then found
to be
\be
F_{A}(t^{2}) &=&-4M_{P}
        \int_{0}^{\xi} \frac{dx\,d^{2}k_{\bot }}{2(2\pi)^{3}\,\xi}
\Bigg\{
 \frac{2(1+2x)(x+x')}{3}\,\frac{m_{q}-x'(m_{Q}-m_{q})k_{\bot }^{2}
        \Theta}{m_{q}^{2}+k_{\bot}^{2}}
\frac{\Phi\left( x',k_{\bot }^{2}\right)}{x'(1-x')} \, \\ && +
\frac{(1+2x)(1-x''+x)}{3}\,\frac{m_{Q}+(1-x'')(m_{Q}-m_{q})k_{\bot
}^{2}
        \Theta }{m_{Q}^{2}+k_{\bot }^{2}}
\frac{\Phi\left( x'',k_{\bot }^{2}\right)}{x''(1-x'')} \Bigg\} \,,  \nonumber \\
F_{V}(t^{2}) &=&-4M_{P}
        \int_{0}^{\xi} \frac{dx\,d^{2}k_{\bot }}{2\left( 2\pi \right) ^{3}\,\xi}
\Bigg\{
 \frac{2(x+x')}{3}\,\frac{(1+2x)m_{q}-A k_{\bot }^{2}
        \Theta }{m_{q}^{2}+k_{\bot }^{2}}
\frac{\Phi\left( x',k_{\bot }^{2}\right)}{x'(1-x')}
                \nonumber \\
&&~~~~ +
\frac{(1-x''+x)}{3}\,\frac{(1+2x)m_{Q}-
        B k_{\bot }^{2}\Theta }{m_{Q}^{2}
        +k_{\bot }^{2}}
\frac{\Phi\left( x'',k_{\bot }^{2}\right)}{x''(1-x'')} \Bigg\}\,, \label{fffv}
\ee
respectively, where
\be
\xi &=&\frac{p^{+}}{q^{+}}=\frac{M_{p}^{2}}{t^2-M_{p}^{2}}\,,
\nn\\ A &=& x'(m_Q-m_q) -2xm_q\,, \nn\\ B &=& -(1-x'')( m_Q-m_q)-2
x m_q\,,\nn\\
\Phi (x,k_{\bot}^2) &=& N\left( {\frac{x(1-x) }{2(M_0^2-(m_Q-m_q)^2)}}%
\right)^{1/2} \sqrt{{\frac{dk_{z}}{dx}}}\exp \left( -{\frac{\vec{k}^{2}}{%
2\omega_M^2}}\right)\,,  \nonumber \\
\Theta &=& {\frac{1}{\Phi(x,k_{\bot}^2) }} {\frac{d\Phi(x,k_{\bot}^{2})}{%
dk_{\bot}^2}} \, ,  \nonumber \\
x^{\prime}&=&x\left(\frac{t^2-M_{p}^{2}}{M_{p}^{2}}\right),\
x''=1-x\left(\frac{t^2-M_{p}^{2}}{M_{p}^{2}}\right),\
\vec{k}=(\vec{k}_{\bot}, \vec{k}_{z}) \,.
\ee

\se{Decay widths}

\ \ \ To compute numerical values of the form factors, we use
$f_{\pi}=0.925$, $m_{Q}=m_{q}=m_{u}=m_{d}=0.25$, $M_{\pi}=0.14$,
$\omega_{\pi} =0.3$ for the $\pi$ meson, and $f_{K}=0.113$,
$m_{Q}=m_{s}=0.4$, $m_{q}=m_{u}=0.25$, $M_{K}=0.495$, $\omega_K
=0.37$ for the $K$ meson in GeV \cite{pdg,fpi}, respectively. We
start with the decay of $\tau \to \nu_{\tau} \pi \gamma$. We
define $x=2l\cdot q/m_{t}^{2}$ and $y=2l\cdot p/m_{t}^{2}$. In the
$\tau$ rest frame, $x\ (y)$ corresponds to the photon (pion )
energy of $E_{\gamma\,(\pi)}$, expressed in units of $m_{\tau}/2$
as \be x=\frac{2E_{\gamma}}{m_{\tau}} \,~~~~~\mbox{and}~~~~~~
y=\frac{2E_{\pi}}{m_{\tau}}\,. %
\ee In terms of $x$ and $y$, one has the following kinematics:
\be
p \cdot q &=& \frac{m_{\tau}^2}{2}(x+y-1-r)\,, \nn\\%
t^2 &=& (s-k)^2 = (p + q)^2 = m_{\tau}^2 (x+y-1)\,. %
\ee
 The physical
allowed regions for
 $x$ and $y$ are given by:
\be
0 & \leq x  \leq & 1-r\,, \nn\\
1-x+\frac{r}{1 -x} & \leq y \leq & 1+r\,, \ee with \be r =
\left(\frac{m_{\pi}}{m_{\tau}} \right)^2 \sim 6.21 \times 10^{-3}
\ee where we have used $m_{\tau}=1.777~~\mbox{GeV}$. We now
calculate the differential decay rate of $\tau^- \to \nu_{\tau}
\pi^- \gamma$, given by \be d\Gamma(\tau \to \nu_{\tau} \pi
\gamma)= \frac{1}{2 m_{\tau}} \delta^{(4)}(l-q-p-k) | {\cal M} |^2
{d\vec{q}\over (2\pi)^3 2E_{\gamma }} {d\vec{p}\over (2\pi)^3
2E_{\pi}} {d\vec{k}\over
(2\pi)^3 2E_{\nu}}\,. %
\ee%
 In the tau rest frame,
 one has
\be
\frac{d^2 \Gamma}{dx\, dy} = \frac{m_{\tau}}{256 \pi^3}
|{\cal M}|^2 \,,
\ee
 where
\be
|{\cal M}|^2=|{\cal M}_{IB}|^2+|{\cal M}_{SD}|^2+2Re({\cal M}_{IB}{\cal
M}_{SD}^{*})\,.
\ee
By writing the decay width $\Gamma$ in terms of the three different source as in Eq.~(38),
i.e.,
\be
\Gamma_{total} & = & \Gamma_{IB} + \Gamma_{SD} + \Gamma_{INT} %
\ee
and the non-radiative decay ($\tau^- \to \nu_\tau \pi^-$) width
\be
\Gamma_{\tau \to \nu_{\tau} \pi} = \frac{G_F^2 |V_{ud}|^2 \theta_c
f_{\pi}^{2}}{8 \pi} m_{\tau}^{3} (1 - r)^2=2.44 \times 10^{-10}
\, \mbox{MeV}\,, %
\ee %
we obtain that
\be
\frac{d^2\Gamma_{IB}}{dx \, dy} & = &
\frac{\alpha}{2 \pi} \rho_{IB}(x,y) \frac{\Gamma_{\tau \to
\nu_{\tau} \pi}}{(1-r)^2}\,,
\nonumber \\
\frac{d^2\Gamma_{SD}}{dx \, dy} & = & \frac{\alpha}{16 \pi}
\frac{m_{\tau}^{4}}{f_{\pi}^{2} m_{\pi}^{2}}
\Bigg[ |F_V|^2 \rho_{VV}(x,y) \nonumber \\
&&+ 2Re(F_V F_A^\star)
\rho_{VA}(x,y)
+ |F_A|^2 \rho_{AA}(x,y) \Bigg] \frac{\Gamma_{\tau \to \nu_{\tau} \pi}}{(1-r)^2}\,,
\nonumber \\
\frac{d^2\Gamma_{INT}}{dx\,dy} & = & \frac{\alpha}{2 \sqrt{2} \pi}
\frac{m_{\tau}^{2}}{f_{\pi} m_{\pi}} \left[
\rho_{INTV}(x,y)\,Re(F_V) + \rho_{INTA}(x,y)\,Re(F_A) \right]
\frac{\Gamma_{\tau \to \nu_{\tau} \pi}}{(1-r)^2}\,,
\label{density} \ee where \be \rho_{IB} (x,y) & = & \frac{[r^2 (x
+ 2) - 2 r (x + y) + (x + y - 1)(x^2 - 3x + 2 + xy)](r - y + 1)}
{(r - x - y +1)^2 x^2}\,, \nonumber \\ \rho_{VV} (x,y) & = & -
[r^2 (x + y) + 2 r (1 - y) (x + y) + (x + y - 1)(x^2+y^2-x- y)]\,,
\nonumber \\ \rho_{AA} (x,y) & = & \rho_{VV}(x,y,r)\,, \nonumber
\\ \rho_{VA}(x,y) & = & [r^2
(x + y) + (1 - x - y)(y-x)] (r^2 - x - y + 1)\,, \nonumber \\
\rho_{INTA}(x,y) & = & - \frac{(r - x - y + 1)(r - y + 1)}{x}\,,
\nonumber \\ \rho_{INTV}(x,y) & = & \frac{[r^2 - 2 r (x + y) + (1
- x + y) (x + y - 1)](r - y + 1)}{(r - x - y + 1) x}\,. %
\ee%
To simplify our calculations, we now introduce $\lambda$ as a new
parameter
% rescale momentum transfer squared and substitute the
%variable $y$:
\be
\lambda=\frac{t^2}{m_{\tau}^{2}}=x+y-1\,.%
 \ee%
  The kinematical
boundaries for $x$ and $\lambda$ are given by:
\be
\lambda - r & \leq x  \leq & 1 - \frac{r}{\lambda} \,, \nonumber
\\ %
r & \leq  \lambda  \leq & 1 \,.%
\ee
 By integrating the variable $x$
in the phase space, from Eq.~(41), we derive the expressions of
differential decay widths for the invariant mass spectrum as:
\be
\frac{d\Gamma_{IB}}{d\lambda} & = &
\frac{\alpha}{2 \pi} \Bigg[ (1 - \lambda) (r^2 + 2 r \lambda - 4 \lambda + \lambda^2)
 \nonumber \\
&& + (r^2 \lambda + 2 r \lambda - 2 \lambda -2 \lambda^2 +
\lambda^3)
\ln \lambda \Bigg]\frac{1}{\lambda^2 - r \lambda}
\frac{\Gamma_{\tau \to \nu_{\tau} \pi}}{(1 - r )^2}\,,
\nonumber \\
\frac{d\Gamma_{SD}}{d\lambda} & = &\frac{\alpha}{48 \pi}
\frac{m_{\tau}^4}{f_{\pi}^{2} m_{\pi}^{2}}
\frac{(\lambda - 1)^2 (\lambda - r)^3 (1 + 2\lambda)}{\lambda^2}
(|F_V|^2+|F_A|^2)\frac{\Gamma_{\tau \to \nu_{\tau} \pi}}{(1 - r)^2}\,,
\nonumber \\
\frac{d\Gamma_{INTV}}{d\lambda} & = &\frac{\alpha}{2 \sqrt{2} \pi}
\frac{m_{\tau}^2}{f_{\pi} m_{\pi}}\frac{(\lambda-r)^2 (1 -\lambda
+ \lambda \ln \lambda)}{\lambda}\,Re(F_V)
\frac{\Gamma_{\tau \to \nu_{\tau} \pi}}{(1 - r)^2}\,,
\nonumber \\
\frac{d\Gamma_{INTA}}{d\lambda} & = &
\frac{\alpha}{2 \sqrt{2} \pi}
\frac{m_{\tau}^2}{f_{\pi} m_{\pi}}
\Bigg[ (1 - \lambda) (r -2 \lambda -1 ) \nonumber \\
& & + (r \lambda - 2\lambda - \lambda^2) \ln \lambda \Bigg]
\frac{\lambda-r}{\lambda}\,Re(F_A)
\frac{\Gamma_{\tau \to \nu_{\tau} \pi}}{(1 - r)^2}\,.%
\label{n45}
\ee%
Note that for the IB part in Eqs.~(\ref{density}) and (\ref{n45})
the contribution is infrared divergent when $x$ closed to $0$ and,
moreover, there is also an enhancement in the limit $\lambda \to
r$. This means that the IB term contains the logarithm divergent
as $t^2 \to m_{\pi}^{2}$. To obtain the decay width of $\tau \to
\nu_{\tau} \pi \gamma$, a cut on the photon energy is needed. The
differential decay width $d\Gamma_{IB}/d\lambda$ as a function of
$\lambda=t^2/m_{\tau}^{2}$ in terms of $\Gamma_{\tau \to
\nu_{\tau} \pi}$ is shown in Figures 2 and 3 with two different
cuts of $E_{\gamma}=50$ and 400 MeV, respectively. In Table 1, we
list the integrated decay width ratio of $R_\pi=\Gamma_{\tau \to
\nu_{\tau} \pi \gamma}/\Gamma_{\tau \to \nu_{\tau} \pi}$ for the
two cuts.
\begin{table}[h]\caption{Integrated Decay width ratio for
$\tau \to \nu_{\tau} \pi \gamma$.}
\begin{center}
\begin{tabular}{|c|c|c|c|c|c|c|}
\hline Integrated decay width ratio& IB & SD & INT & Sum &
Ref.\cite{tau2} & Ref.\cite{tau3} \\ \hline \hline
$10^{-3}R_\pi
%\Gamma(\tau \to \nu_{\tau} \pi \gamma)
(E_{\gamma}\geq 50)$ MeV & $13.1$ & $1.48$ & $1.61$ & $16.2$ & $-$
& $14.8$\\ \hline $10^{-3}R_\pi
% \Gamma(\tau \to \nu_{\tau} \pi\gamma)
(E_{\gamma}\geq 400)$ MeV & $1.48$ & $1.48$ & $0.90$ & $3.86$ &
$1.16$ & $2.76$\\ \hline
\end{tabular}\end{center}\end{table}
%We now compare our results with those in Ref. \cite{tau3}.
As
shown in Table 1, our results for the decay width of
$\tau \to \nu_{\tau} \pi \gamma$ are larger than those in Refs. \cite{tau2} and

\cite{tau3}, respectively, due to our bigger SD contribution.
From Table 1, it is interesting to see that $\Gamma_{INT}$ which
depends on the photon energy cut can be as large as $\Gamma_{SD}$.

For the decay of $\tau \to \nu_{\tau} K \gamma$, we use
$\Gamma_{\tau \to \nu_{\tau} K } =1.759\times 10^{-11}$ MeV. The
differential decays width for $\Lambda=t^2/m_\tau^2\geq 0.2\ GeV$
in terms of $\Gamma_{\tau \to \nu_{\tau} K }$ is plotted in
Figure~4. From the figure, we see that the SD contribution to the
decay width is dominant for  values of $t^2$. The reason is that
the large kaon mass suppresses the IB contribution. In Table 2, we
show the decay width ratio of $R_K=\Gamma_{\tau \to \nu_{\tau} K
\gamma}/\Gamma_{\tau \to \nu_{\tau} K}$
%$\tau \to \nu_{\tau} K\gamma$
with two different invariant mass cuts of $t=800$ and
$1200$ MeV.
\begin{table}[h]\caption{Integrated decay width ratio for
$\tau \to \nu_{\tau} K \gamma$.}
\begin{center}
\begin{tabular}{|c|c|c|c|c|c|}
\hline Integrated Decay width ratio& IB & SD & INT & Sum  &
Ref.\cite{tau3} \\ \hline \hline $10^{-4}R_K
%\Gamma(\tau \to \nu \pi\gamma)
(t \geq 800 $ MeV ) & $6.51$ & $7.36$ & $5.26$ & $19.13$ &
$35.8$\\ \hline $10^{-4}R_K
%\Gamma(\tau \to \nu \pi \gamma)
(t \geq 1200 $ MeV ) & $0.72$ & $3.80$ & $0.86$ & $5.38$ & $9.1$\\
\hline
\end{tabular}\end{center}\end{table}
Form Table 2, we find that the results in Ref. \cite{tau3} are
twice as large as our predictions.

We now study the ratio
\be
R={\sum_{P=\pi,\,K}Br(\tau \to \nu_{\tau} P)\over Br(\tau \to e \nu_e
\bar{\nu}_\tau)}%
 \ee
 and examine both theoretical and experimental values.
In the standard model, the ratio is given by:
\be
R^{SM}=0.646\,,
 \ee while the recent experimental average is
\cite{pdg}
\be
R^{exp}=0.66100 \pm 0.00725\,.
\label{n48}
 \ee%
  Clearly, there is a discrepancy
between $R^{SM}$ and $R^{exp}$. However, it is believed that it
arises from the radiative corrections.
 At $O(\alpha)$,
the radiative corrected decay width is found to be
\be
\Gamma(\tau^- \to \nu_{\tau} \pi^- (\gamma))=\Gamma(\tau^- \to \nu_{\tau} \pi^-
)+\Gamma(\tau^- \to \nu_{\tau} \pi^- \gamma) \sim 2.48 \times 10^{-10}~~
\mbox{MeV} \, %
\label{n49}
\ee %
with $E_{\gamma} \geq 50\, \mbox{MeV}$.
 Similarly, for the mode
with $K$, one has that
\be
\Gamma(\tau^- \to \nu_{\tau} K^- (\gamma)) \sim 1.61 \times 10^{-11}~~
\mbox{MeV} \, %
\ee with the same photon energy cut as the pion mode.
%$E_{\gamma}\geq 50\, \mbox{MeV}$.
In the standard model, the radiative
corrected width of $\tau \to e \nu \bar{\nu}$ \cite{mar} is given
by
\be
\Gamma(\tau^- \to e^- \nu_e \bar{\nu}_{\tau} (\gamma))&=&
%\frac{G_{F}^{2} m_{\tau}^{5}}{192 \pi^{3}} f\bigg(\frac{m_{e}^{2}}{m_{\tau}^{2}}\bigg)
%\bigg(1+\frac{3}{5} \frac{m_{\tau}^{2}}{m_{W}^{2}}\bigg)
%\bigg[1+\frac{\alpha}{2 \pi} (\frac{25}{4}-\pi^2)\bigg] \nonumber \\
%&=&
(4.033 \pm 0.005)\times 10^{-10}~~ \mbox{MeV} \,.%
\label{n51}%
 \ee %
% where
%\be
%f(x)=1-8x+8x^3-x^4-12x^2\ln x\,.%
% \ee%
  From the Eq. (\ref{n49})-(\ref{n51}), we obtain
\be R^{Theor,\, O(\alpha)}&\sim&0.655 \ee which agrees with
experimental data in Eq.~(\ref{n48}) within the errors.

\se{Conclusions}

\ \ \

We have studied the decays of $\tau \to \nu_{\tau} \pi(K) \gamma$
in the light front quark model. We have calculated the form
factors and used them to evaluate the decay widths. We have found
that, in the standard model, the decay widths are $1.62\times
10^{-2}\,\, (3.86\times 10^{-3})\, \Gamma_{\tau \to \nu_{\tau} \pi
}$ and $1.91\times 10^{-3}\,\,(5.38\times 10^{-4})\, \Gamma_{\tau
\to \nu_{\tau} K }$ with the cuts of
$E_{\gamma}=50\,\,(E_{\gamma}=400)$ MeV and $t_{0}= 800\,\,(1200)$
MeV for $\tau \to \nu_{\tau} \pi \gamma$ and $\tau \to \nu_{\tau}
K \gamma$, respectively. We have also shown that with including
the radiative decay widths, the experimental rate for $\tau\to
\nu_\tau (\pi,K)$ can be understood. In future, as the tau-charm
factories will produce a large number of samples of $\tau$ lepton
pairs, these rare $\tau$ decays should be precisely measured and
thus the structure-dependent form-factors can be well tested. In
this case, the decays can be used to probe new physics similar to
$K_{l3}$ \cite{gengK0} and $K_{l2\gamma}$ decays \cite{gengK}.\\

\noindent
{\bf Acknowledgments}

This work was supported in part by the National Science Council of the
Republic of China under the Grant Nos. NSC91-2112-M-007-043 and
NSC91-2112-M-007-046.

\newpage

\begin{figure}[h]

\vskip 12cm
\begin{center}

\begin{picture}(100,100)(100,0)

\Photon(50,320)(100,320){5}{4}

\ArrowLine(100,320)(150,370)

\ArrowLine(150,370)(200,320)
\ArrowLine(100,320)(200,320)
\Photon(150,370)(180,400){5}{4}
\Line(200,319)(250,319)
\Line(200,321)(250,321)
\Vertex(200,320){3}
\Text(115,360)[l]{$k_{1}$}
\Text(175,360)[l]{$k_{2},\,k^{'}_{2}$}
\Text(160,305)[r]{$k^{'}_{1}$}
\Text(70,305)[r]{$t$}
\Text(240,305)[r]{$\pi (p)$}
\Text(160,400)[r]{$\gamma (q)$}
\Text(160,260)[r]{$\quad\mbox{(a)}$}

\Photon(50,100)(100,100){5}{4}

\ArrowLine(100,100)(150,150)

\ArrowLine(150,150)(200,100)
\ArrowLine(100,100)(200,100)
\Photon(150,150)(180,180){5}{4}
\Line(200,99)(250,99)
\Line(200,101)(250,101)
\Vertex(200,100){3}
\Text(115,140)[l]{$k_{1}$}
\Text(175,140)[l]{$k_{2},\,k^{'}_{1}$}
\Text(160,85)[r]{$k^{'}_{2}$}
\Text(70,85)[r]{$t$}
\Text(240,85)[r]{$\pi (p)$}
\Text(160,180)[r]{$\gamma (q)$}
\Text(160,40)[r]{$\quad\mbox{(b)}$}
\end{picture}
\vskip 0.5cm
\caption{ Loop diagrams that contribute $\tau \to
\nu_{\tau} \pi \gamma$.
 }

\end{center}

\end{figure}

\newpage
\begin{figure}[h]
\includegraphics{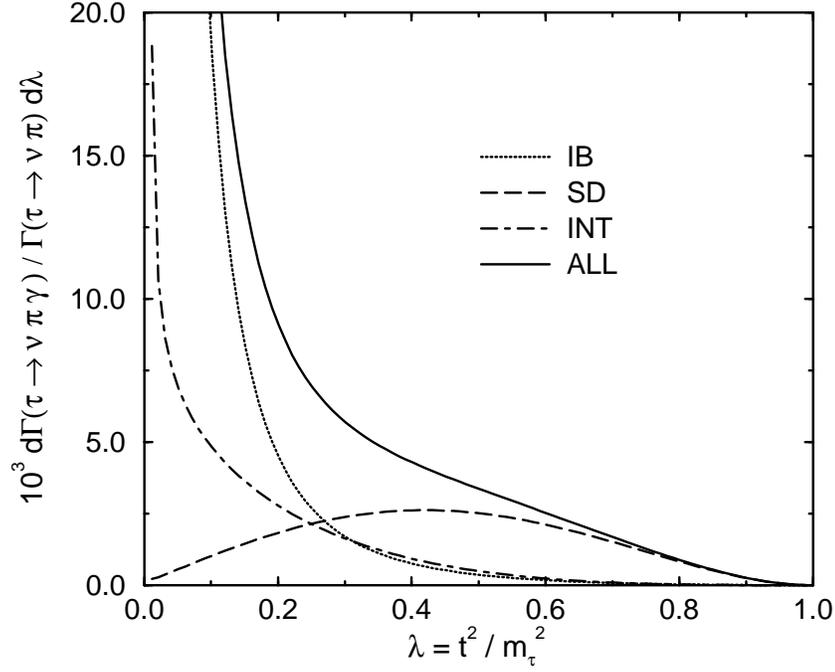} \vskip 13cm \caption{ Differential decay width of
$\Gamma(\tau \to \nu_{\tau} \pi \gamma)$ in terms of $\Gamma_{\tau
\to \nu_{\tau} \pi}$ as a function of $\lambda=t^2/m_{\tau}^2$
with $E_{\gamma} \geq 50$ MeV. The dot, dash, dash-dot and solid
curves stand for the contributions of IB, SD, INT, and total
parts, respectively.}
\end{figure}

\newpage
\begin{figure}[h]
\includegraphics{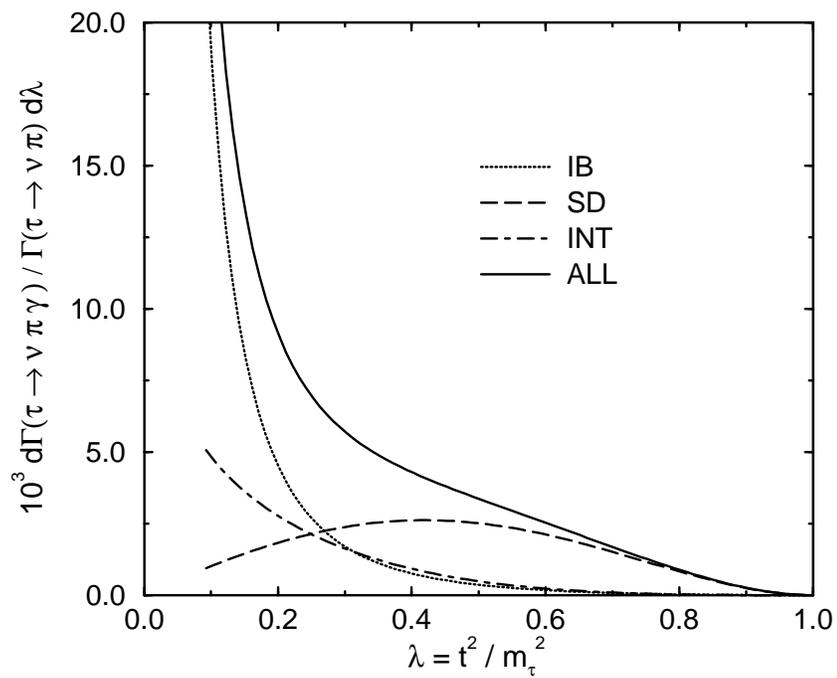} \vskip 13cm
 \caption{ Same as Figure 2 but with
$E_{\gamma} \geq 400$ MeV.}
\end{figure}

\newpage
\begin{figure}[h]
\includegraphics{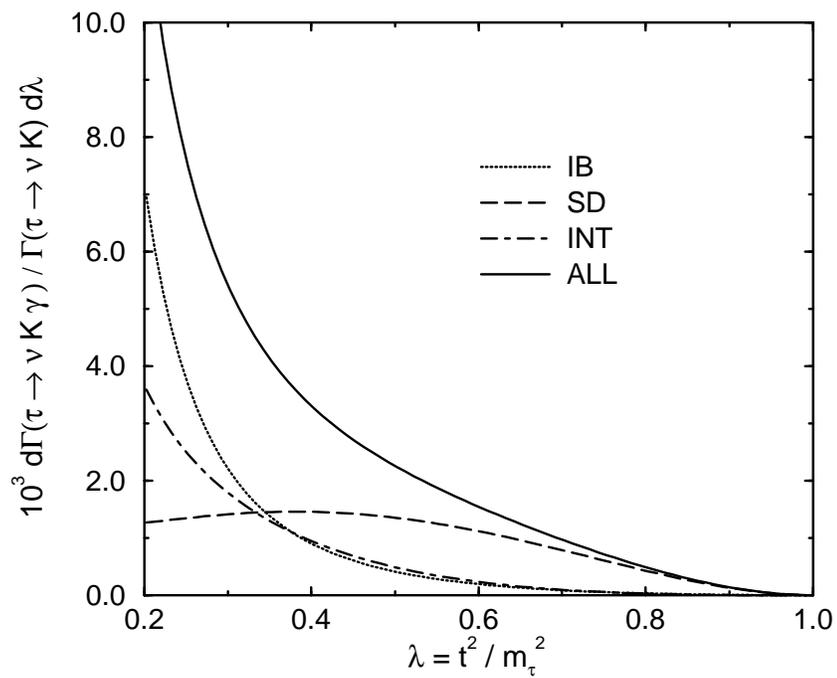} \vskip 13cm \caption{ Differential decay width of
$\Gamma(\tau \to \nu_{\tau} K \gamma)$ as a function of
$\lambda=t^2/m_{\tau}^{2}$ in terms of $\Gamma_{\tau \to
\nu_{\tau} K}$. Caption is the same as in Figure 2.}
\end{figure}

\end{document}